\documentclass[epj]{webofc}
\usepackage[varg]{txfonts}   
\usepackage{graphicx,color} 
\usepackage{bm}       
\usepackage{amsmath}  
\usepackage{amssymb}
\usepackage{float,mathtools,amsfonts,slashed}
\woctitle{21st International Conference on Few-Body Problems in Physics}
\begin{document}
\title{A systematic approach to sketch Bethe-Salpeter equation}
\author{Si-xue Qin\inst{1}\fnsep\thanks{\email{sqin@anl.gov}} 
}

\institute{Physics Division, Argonne National Laboratory, Argonne, Illinois 60439, USA
          }

\abstract{
To study meson properties, one needs to solve the gap equation for the quark propagator and the Bethe-Salpeter (BS) equation for the meson wavefunction, self-consistently. The gluon propagator, the quark-gluon vertex, and the quark--anti-quark scattering kernel are key pieces to solve those equations. Predicted by lattice-QCD and Dyson-Schwinger analyses of QCD’s gauge sector, gluons are non-perturbatively massive. In the matter sector, the modeled gluon propagator which can produce a veracious description of meson properties needs to possess a mass scale, accordingly. Solving the well-known longitudinal Ward-Green-Takahashi identities (WGTIs) and the less-known transverse counterparts together, one obtains a nontrivial solution which can shed light on the structure of the quark-gluon vertex. It is highlighted that the phenomenologically proposed anomalous chromomagnetic moment (ACM) vertex originates from the QCD Lagrangian symmetries and its strength is proportional to the magnitude of dynamical chiral symmetry breaking (DCSB). The color-singlet vector and axial-vector WGTIs can relate the BS kernel and the dressed quark-gluon vertex to each other. Using the relation, one can truncate the gap equation and the BS equation, systematically, without violating crucial symmetries, e.g., gauge symmetry and chiral symmetry.
}
\maketitle
\section{Introduction}
The visible mass of the universe is mainly contributed by hadrons -- bound states of fundamental blocks, i.e., quarks and gluons. Their dynamics is described by quantum chromodynamics (QCD) -- the strong interaction sector of the Standard Model. QCD has two fascinating features: dynamical chiral symmetry breaking (DCSB) and confinement, which are not apparent in its Lagrangian. It is believed that DCSB is responsible for the generation of mass from nothing (the mass of constituent light quarks is several hundreds MeV), and thus the Higgs mechanism has almost nothing to do with the origin of the mass of the visible matter (the mass of current light quarks is only several MeV). On the other hand, the fundamental degree of freedoms, i.e., quarks and gluons, are confined and cannot directly be detected. To understand confinement is one of the greatest challenges in modern physics. Neither DCSB nor confinement is understood perturbatively. The key to the whole story is to solve QCD non-perturbatively. 

In order to solve QCD non-perturbatively, we try to study its equations of motion, i.e., Dyson-Schwinger equations (DSEs) \cite{Roberts:1994dr}. Specifically, we need to deal with three categories of equations, i.e., the one-body gap equation for quarks, the two-body Bethe-Salpeter (BS) equation for mesons, and the three-body Faddeev equation for baryons. In the rest of this paper, we will focus on the gap equation and the BS equation, i.e.,
\begin{eqnarray}
\parbox{53mm}{\includegraphics[width=\linewidth]{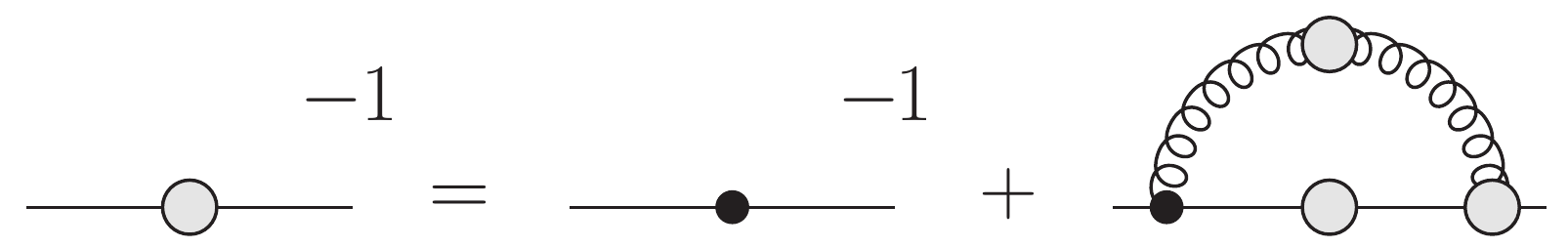}} \text{\quad and \quad} \parbox{50mm}{\includegraphics[width=\linewidth]{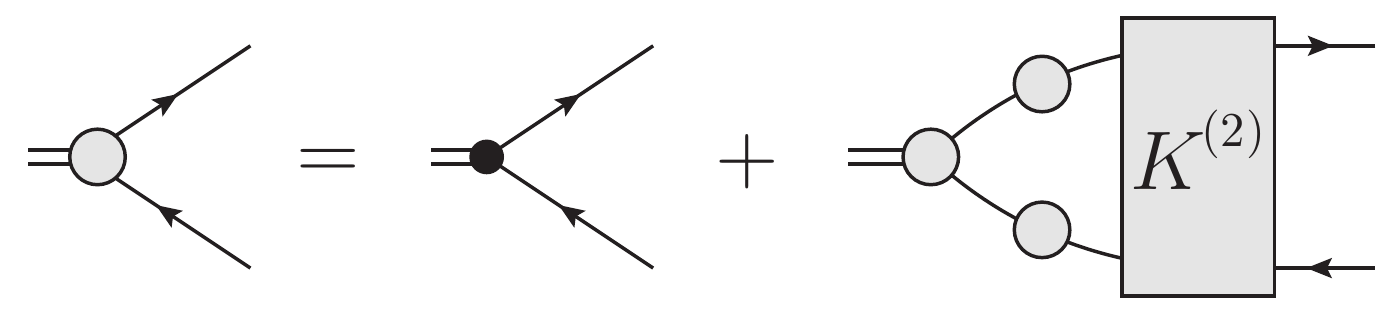}}\,,
\end{eqnarray}
where gray circular blobs denote dressed propagators and 
vertices, $K^{(2)}$ denotes the full quark--anti-quark 
scattering kernel, and black dots denote bare propagators or vertices. Once the gluon propagator, the quark-gluon vertex, and the scattering kernel are specified, the two equations can be solved, and thus meson properties can be studied accordingly.

\section{Gluon propagator}
In pure SU(3) Yang-Mills theory, the Lagrangian is gauge invariant and thus does not include a mass scale. Certainly, for the standard perturbative solution of QCD, the gluon is massless. However, as discussed in Ref. \cite{Cornwall:1981zr}, the gluon can be non-perturbatively massive. 

In Euclidean space and Landau-gauge, the gluon propagator is decomposed as
\begin{eqnarray}
	\mathcal{D}_{\mu\nu}(k) = \left( \delta_{\mu\nu} - \frac{k_\mu k_\nu}{k^2} \right) D(k^2) \,, \quad D(k^2) = \frac{Z(k^2)}{k^2 + m_g^2(k^2)} \,,
\end{eqnarray}
where $D(k^2)$ is the dressing function of gluon and has a massive-type form according to lattice QCD \cite{Oliveira:2010xc} or DSE \cite{Aguilar:2002tc} calculations; $Z(k^2)$ is the wavefunction renormalization function, and $m_g(k^2)$ is the running gluon mass. Inspired by the result in one-loop perturbative QCD, 
 one can parameterize $Z(k^2)$ and $m_g(k^2)$ as
\begin{eqnarray}
	Z(k^2) = z_0 \left( \ln \frac{k^2 + r M_g^2}{\Lambda^2} \right)^{-\gamma},\quad m_g^2(k^2) = \frac{M_g^4}{k^2+M_g^2} \,,
\end{eqnarray} 
where $\gamma$ is the gluon anomalous dimension; $z_0$ , $r$, $\Lambda$, and $M_g$ are the fitting parameters. Using such a form, lattice data can be well fitted with a typical gluon mass scale $M_g \sim 700$ MeV \cite{Oliveira:2010xc}.

Putting the non-perturbatively massive gluon into consideration, we proposed a realistic interaction model \cite{Qin:2011dd} as
\begin{eqnarray}
	g^2 \mathcal{D}_{\mu\nu}(k) = k^2 \mathcal{G}(k^2) \mathcal{D}^{\rm free}_{\mu\nu}(k)  = [k^2\mathcal{G}_{\rm IR}(k^2) + \tilde\alpha_{\rm pQCD}(k^2) ]\mathcal{D}^{\rm free}_{\mu\nu}(k) \,,
\end{eqnarray}
wherein $\mathcal{D}^{\rm free}_{\mu\nu}(k)$ is the free-gauge-boson propagator, $\mathcal{G}_{\rm IR}(k^2)$ is an ansatz which dominates the infrared interaction, and $\tilde\alpha_{\rm pQCD}(k^2)$ is a regular continuation of the perturbative-QCD running coupling constant. Explicitly, 
\begin{eqnarray}
	\mathcal{G}(s) = \frac{8\pi^2}{\omega^4} D e^{-s/\omega^2} + \frac{8\pi^2\gamma_m \mathcal{F}(s)}{\ln \left[ \tau + (1 + s/\Lambda_{\rm QCD}^2 )^2 \right]} \approx \frac{4\pi \alpha(s)}{s + m_g^2(s)} \,,
\end{eqnarray} 
where $\gamma_m = 12/25$, $\Lambda_{\rm QCD} =0.234$ GeV; $\tau = e^2 -1$; $\mathcal{F}(s) = [1-\exp(-s)]/s$; and $\alpha(s)$ is the effective running coupling constant. In such a form, the one-loop renormalization-group behavior of QCD is preserved. The two parameters, i.e., the interaction strength $D$ and the interaction width $\omega$, shape the infrared interaction and may implicitly represent higher-order ultraviolet contributions. With the product $D\omega$ fixed, one can obtain a uniformly good description of pseudoscalar and vector mesons \cite{Qin:2011dd}. In the favorable parameter space, the values of $M_g$ are typical. The infrared value of the coupling is, typically, $\alpha_{\rm RL}(0)/\pi \sim 10$ and $\alpha_{\rm DB}(0)/\pi \sim 2$, where RL stands for the simplest rainbow-ladder approximation and DB for the DCSB-improved approximation.

\section{Quark-gluon vertex}
In perturbation theory, the quark-gluon vertex can be calculated order by order in loop expansion. However, since QCD is non-perturbative, the dynamics dressing effect fundamentally alters the appearance of the vertex. In order to expose the structure of the quark-gluon vertex (or the fermion–gauge-boson vertex in general), we solve the familiar longitudinal WGTI and its less well known transverse counterparts together, where we work in the Abelian approximation to avoid dealing with ghost fields.

As a consequence of gauge invariance, the divergence of the dressed-fermion–gauge-boson vertex can be related to the dressed-fermion propagators ($q=k-p$ and $S(p) = 1/[i\gamma\cdot p A(p^2) + B(p^2)]$)
\begin{eqnarray}
	i q_\mu \Gamma_\mu(k,p) = S^{-1}(k) - S^{-1}(p)\,.
\end{eqnarray}

Combining the Lorentz transformation with the gauge and chiral transformations, one can derive the vector and axial-vector transverse WGTIs $(t=k+p)$, respectively:
\begin{eqnarray}
\lefteqn{\hspace*{-0.5cm}q_\mu \Gamma_\nu(k,p)-q_\nu \Gamma_\mu(k,p) =
S^{-1}(p)\sigma_{\mu\nu} + \sigma_{\mu\nu} S^{-1}(k) + 2 i m \Gamma_{\mu\nu}(k,p) + t_\lambda \varepsilon_{\lambda\mu\nu\rho} \Gamma^A_{\rho}(k,p) + A^{V}_{\mu\nu}(k,p)\,,}
\label{eqTWGTI}\\
\lefteqn{\hspace*{-0.5cm}q_\mu \Gamma^A_{\nu}(k,p) - q_\nu \Gamma^A_{\mu}(k,p)=
S^{-1}(p)\sigma^5_{\mu\nu} -\sigma^5_{\mu\nu}S^{-1}(k) + t_\lambda\varepsilon_{\lambda\mu\nu\rho} \Gamma_\rho(k,p) + V^{A}_{\mu\nu}(k,p)\,,}
\label{eqTAWGTI}
\end{eqnarray}
where $m$ is the fermion mass; tr$[\gamma_5\gamma_\mu\gamma_\nu\gamma_\rho\gamma_\sigma]=-4 \varepsilon_{\mu\nu\rho\sigma}$; $\sigma_{\mu\nu}=(i/2)[\gamma_\mu,\gamma_\nu]$, $\sigma^5_{\mu\nu} = \gamma_5\sigma_{\mu\nu}$; $\Gamma_{\mu\nu}(k,p)$ is an inhomogeneous tensor vertex; and $A^V_{\mu\nu}(k,p)$ and $V^A_{\mu\nu}(k,p)$ stands for contributions from high-order Green functions. The transverse WGTIs express curls of the vertices. 

A first observation for those transverse WGTIs is that the identities for different vertices are coupled together. However, with the well-defined projection tensors \cite{Qin:2013mta}
\begin{eqnarray}
T^1_{\mu\nu} = \frac{1}{2} \,  \varepsilon_{\alpha\mu\nu\beta} t_\alpha q_\beta \mathbf{I}_{\rm D}\,,\;
T^2_{\mu\nu} = \frac{1}{2} \, \varepsilon_{\alpha\mu\nu\beta} \gamma_\alpha q_\beta \,,
\end{eqnarray}
one can decouple the identities and isolate the transverse WGTIs for the vector vertex as
\begin{eqnarray}
q\cdot t\, t \cdot \Gamma(k,p) &=& T^1_{\mu\nu} \left[S^{-1}(p)\sigma^5_{\mu\nu} - \sigma^5_{\mu\nu}S^{-1}(k)\right]  + \, t^2  q\cdot \Gamma(k,p) + T^1_{\mu\nu} V^{A}_{\mu\nu}(k,p) \,,
\label{eqtransWTI1}\\
q\cdot t\, \gamma \cdot \Gamma(k,p) &=& T^2_{\mu\nu} \left[S^{-1}(p)\sigma^5_{\mu\nu} - \sigma^5_{\mu\nu}S^{-1}(k)\right] + \, \gamma\cdot t \, q\cdot \Gamma(k,p) + T^2_{\mu\nu} V^{A}_{\mu\nu}(k,p) \,.
\label{eqtransWTI2}
\end{eqnarray}
Choosing a proper basis for the matrix-valued equations, one can obtain a unique solution for the vector vertex. The solution confirms the longitudinal Ball-Chiu vertex \cite{Ball:1980ay} as a precise piece of the full fermion–gauge-boson vertex and exposes the transverse ACM terms, e.g.,
\begin{eqnarray}
	\Gamma^{\rm ACM5}_\mu(k,p) = - \Delta_B(k^2,p^2) \sigma_{\mu\nu}q_\nu \,, \quad \Delta_B(k^2,p^2) = [B(k^2)-B(p^2)]/[k^2-p^2] \,,
\end{eqnarray}
which is comparable with that in Ref. \cite{Chang:2010hb}. Apparently, the ACM strength is proportional to the magnitude of DCSB. Thus, it is expected that the ACM terms have significant consequences in observables.

\section{Scattering kernel}
It has been shown that, in the bare vertex approximation, the one-gluon exchange form for the BS scattering kernel fails to produce meson spectrum above 1 GeV, e.g., $a_1$, $b_1$ meson, radial excitation states of pion and rho meson \cite{Qin:2011xq,Chang:2011ei} . An exact kernel must be valid when the quark-gluon vertex is fully dressed.

The solutions of the inhomogeneous BS equations must satisfy the color-singlet vector and axial-vector WGTIs ($k_\pm = k\pm P$), respectively,
\begin{eqnarray}
	iP_\mu \Gamma_{\mu}(k,P) &=& S^{-1}(k_+)- S^{-1}(k_-) \,,\\
	P_\mu \Gamma_{5\mu}(k,P) + 2im \Gamma_{5}(k,P) &=&  S^{-1}(k_+)i\gamma_5 + i\gamma_5 S^{-1}(k_-) \,.
\end{eqnarray}
The vector WGTI guarantees current conservation, e.g., the pion form factor $F_\pi(Q^2=0) = 1$; and the axial-vector counterpart guarantees pion's twofold role: pseudo-scalar bound-state and Goldstone boson, i.e., pion must be massless in the chiral limit \cite{Chang:2009zb,Qin:2014vya}. A scattering kernel which is useful for studying meson properties must preserve such identities. 

Inserting the inhomogeneous BS equations and the quark gap equation into the WGTIs, one can relate the scattering kernel to the quark-gluon vertex as
\begin{equation}
\smallint_q {K^{(2)}}_{\alpha\alpha',\beta'\beta}[S(q_-)
- S(q_+)]_{\alpha'\beta'} = \smallint_q D_{\mu\nu}(k-q)\gamma_\mu [S(q_+) \Gamma_\nu(q_+,k_+)-S(q_-)\Gamma_\nu(q_-,k_-)]\,,
\end{equation}
\begin{equation}
\smallint_q {K^{(2)}}_{\alpha\alpha',\beta'\beta} [S(q_+)\gamma_5 + \gamma_5 S(q_-)]_{\alpha'\beta'} = \smallint_q D_{\mu\nu}(k-q)
\gamma_\mu [S(q_+) \Gamma_\nu(q_+,k_+)\gamma_5 - \gamma_5 S(q_-)
\Gamma_\nu(q_-,k_-)] \,.
\end{equation}
Using the relation and a proper ansatz, one can construct the scattering kernel for a specified vertex, and vice versa.

\section{Epilogue}
This paper explained a systematic approach to sketch BS equation: modeling the gluon propagator, investigating the structure of the quark-gluon vertex, and constructing the quark--anti-quark scattering kernel. Using this approach, the spectrum of light-flavor mesons including ground and radial excitation states can be well described. Moreover, the approach can potentially be extended to more applications, e.g., meson form factors and baryon properties in the diquark picture.

\begin{acknowledgement}
The author would like to thank C. D. Roberts for helpful discussions. The work was supported by the Office of the Director at Argonne National Laboratory through the Named Postdoctoral Fellowship Program -- Maria Goeppert Mayer Fellowship.
\end{acknowledgement}

%
%
%
%
%


\end{document}